\documentclass[pdflatex,sn-nature]{sn-jnl}

\usepackage{graphicx}%
\usepackage{multirow}%
\usepackage{amsmath,amssymb,amsfonts}%
\usepackage{amsthm}%
\usepackage{mathrsfs}%
\usepackage[title]{appendix}%
\usepackage{xcolor}%
\usepackage{textcomp}%
\usepackage{manyfoot}%
\usepackage{color}
\usepackage{bbm}
\usepackage{bm}
\usepackage{booktabs}%
\usepackage{algorithm}%
\usepackage{algorithmicx}%
\usepackage{wrapfig}
\usepackage{algpseudocode}%
\usepackage{listings}%
\usepackage{tikz}
\usetikzlibrary{patterns,shapes,calc}
\usepackage{braket}
\usepackage{float}
\usepackage[acronym]{glossaries}

\makeglossaries
\newacronym{bz}{BZ}{Brillouin zone}
\newacronym{ml}{ML}{machine learning}
\newacronym{sota}{SOTA}{state-of-the-art}
\newacronym{nn}{NN}{neural network}
\newacronym{ann}{ANN}{Artificial Neural Networks}
\newacronym{mlp}{MLP}{Many Layer Perceptron}
\newacronym{nnn}{NNN}{Naive Neural Network}
\newacronym{csnn}{CSNN}{crystal set neural network}
\newacronym{cgnn}{CGNN}{crystal graph neural network}
\newacronym{ccnn}{CCNN}{crystal convolution neural network}
\newacronym{cann}{CANN}{crystal attention neural network}
\newacronym{dft}{DFT}{density functional theory}
\newacronym{tm}{TM}{topological materials}
\newacronym{tqc}{TQC}{topological quantum chemistry}
\newacronym{tb}{TB}{tight binding}
\newacronym{ebr}{EBR}{elementary band representation}
\newacronym{gbt}{GBT}{gradient boosted trees}
\newacronym{rmse}{RMSE}{root mean squared error}
\newacronym{mae}{MAE}{mean absolute error}

\newacronym{trm}{tM}{trivial material}
\newacronym{lcebr}{LCEBR}{linear combination of elementary band representations}
\newacronym{ti}{TI}{topological insulator}
\newacronym{nlc}{NLC}{no linear combination}
\newacronym{sebr}{SEBR}{split elementary band representation}
\newacronym{tsm}{TSM}{topological semimetal}
\newacronym{es}{ES}{enforced semimetal}
\newacronym{esfd}{ESFD}{enforced semimetal with Fermi degeneracy}

\theoremstyle{thmstyleone}%
\theoremstyle{thmstyletwo}%

\theoremstyle{thmstylethree}%

\begin{document}

\title{Faithful novel machine learning for predicting quantum properties} 

\author[1]{\fnm{Gavin} \sur{Nop}}

\author[2]{\fnm{Micah} \sur{Mundy}}

\author[1]{\fnm{Jonathan D. H.} \sur{Smith}}

\author[3]{\fnm{Durga} \sur{Paudyal}}

\affil[1]{\orgdiv{Department of Mathematics}, \orgname{Iowa State University}, \orgaddress{\city{Ames}, \postcode{50011}, \state{Iowa}, \country{USA}}}

\affil[2]{\orgdiv{Department of Mechanical Engineering}, \orgname{Iowa State University}, \orgaddress{\city{Ames}, \postcode{50011}, \state{Iowa}, \country{USA}}}

\affil[3]{\orgdiv{Department of Physics and Astronomy}, \orgname{University of Iowa}, \orgaddress{\city{Iowa City}, \postcode{52242}, \state{Iowa}, \country{USA}}}


\date{\today}

\abstract{
\textcolor{black}{
Machine learning (ML)
}
has accelerated the process of materials classification, particularly with \gls{cgnn} architectures. 
Advanced deep networks have hitherto proved challenging to build and train for the task of materials classification. Here, by adoption of faithful representations, we refine current  
\textcolor{black}{
ML
}
networks and effectively implement advanced deep networks for generic crystalline quantum materials property prediction and optimization. 
The new models achieve strong performance in predicting topological properties, magnetic properties, formation energies, and symmetry groups, and the \gls{cgnn} generates state-of-the-art predictions for \gls{tqc} materials. Full implementations and automated methods for data handling and materials predictions are provided, facilitating the use of deep \textcolor{black}{ML} methods in quantum materials science. 
}

\maketitle

\section{Introduction}

Quantum materials classification and regression occupy a crucial space in the identification of new materials and the optimization of their properties to facilitate novel energy and quantum technology solutions. The primary tools for modern materials exploration, such as \gls{dft}, often require days, weeks, or even months to compute  properties of complex materials such as topological indices, micromagnetic inputs, and electronic structure parameters \cite{vasp}.  \gls{ml} is becoming an increasingly prevalent tool for materials research, given the availability of the Materials Project and other online datasets, with an influx of publications on \gls{ml} designs and applications \cite{lotsofML1, lotsofML2, lotsofML3}. Many of these methods are highly dependent on the materials properties being examined. They may be difficult to implement and bring to convergence. Thus, there is a clear need for the development of a reliable, fast approach to model and correlate diverse materials properties.

As most foundational \gls{ml} models require fixed tensor dimensions for input, early uses of \gls{ml} algorithms for materials research typically hashed properties of the atoms in the primitive cell to produce predictions with random forests, simple multilayer perceptrons, and other techniques \cite{schmidt2019recent}. Recent developments build on this with \gls{cgnn} and convolutional networks to achieve better results \cite{ml-cgnn, zheng2018machine}. \gls{ml} algorithms show great promise, even for predicting non-trivial topological indices \cite{sun2018deep}.

Here we develop four fully general \gls{ml} algorithms which can predict and categorize arbitrary material properties. Each model employs a faithful embedding of the underlying materials, and is fully capable of distinguishing any pair of unique materials, side-stepping the representational reduction employed by many current models. This enables immediate prediction of \textit{any} physical phenomenon with minimal \gls{ml} experience. Our models are tested primarily on the topological data enabled by \gls{tqc} \cite{tqc-dataset}. Further, all models achieve exemplary performance with purely structural information about the materials being modeled, not requiring additional experimental data. Formation energy and magnetic ordering are tested to demonstrate the ability of the models to adapt to arbitrary settings. State-of-the-art is achieved for \gls{tqc} classification. Implementations and pre-trained models are provided in \texttt{GitHub}. 
As the majority of time in \gls{ml} is spent on dataset verification, further tools are provided to automatically extract and purify new materials datasets. This development transforms \gls{ml} from a demonstrative technology in materials science to a tool that is readily available for experimental and theoretical materials researchers. 

\section{Theoretical formulation and approaches}\label{s:theory}

Gadolinium (III) sesquioxide (Gd$_2$O$_3$) with space group 164 is taken as the basic example to illustrate both the physics and the algorithmic processes in the paper. This is a lower symmetry phase than the cubic phase of Gd$_2$O$_3$, and is referenced as material 20470 in the provided \texttt{GitHub} dataset. The trigonal symmetry and smaller number of distinct atoms in the primitive cell are pedagogically and representationally more useful. The relevant formulaic aspects of Gd$_2$O$_3$ for \gls{ml} are the space group, the formation energy, the magnetic classification, and the topological designation, which are 164, -3.723 eV/atom, ferromagnetic, and a split elementary band represented topological insulator, respectively.

\subsection{ Background}\label{s:theory:materials}

Crystalline materials defined by a real space primitive cell were taken as inputs to the \gls{ml} models. Four characteristics are considered for each crystal: the formation energy per atom, the space group ($219$ labels), magnetic (non-magnetic, ferromagnetic, ferrimagnetic, and antiferromagnetic), and topological classifications.

\subsubsection{Crystallographic and band topologies}

Topological indices are non-local over the Brillioun zone (which can still be computed by standard \gls{ml}), and are defined in \gls{tqc} by the following categories and subcategories:
\begin{itemize}
\item
\gls{trm}, which is a \gls{lcebr};
\item
\gls{ti}, labeled as having \gls{nlc}, or as a \gls{sebr}, and 
\item
\gls{tsm}, labeled as an \gls{es}, or as an \gls{esfd}. 
\end{itemize}
Relative to formation energy and magnetic classification, topological state classification is more complex in terms of mathematical and physical origins \cite{tqc-dataset}, involving symmetry-enforced electronic states.

A general program for classification of \gls{ti}'s by symmetry introduced by Zak \cite{topo-history,zak1981band} relied on band representations. This program culminated in the enumeration of all possible trivial band representations, resulting in a predicted 2D and 3D $26,938$ \gls{tm}s via \gls{tqc} \cite{slager2013space, po2017symmetry, tqc-dataset}. The resulting dataset was curated to train an \gls{ml} model achieving an accuracy of 
86\% (as compared to the baseline accuracy of 
50\% by simply marking every material as non-topological) \cite{tqc,tqc-learning, topo-history, topo-generic, topo-symm}. To understand the issues underlying \gls{ml} predictions of these materials, a brief overview of the theory is provided: first formally defining topological insulators, and then providing a framework for understanding \gls{tqc}.

Three primary categories of the \gls{ti} concept are distinguished:
\begin{itemize}
\item[\gls{ti}$_e$]
| determined by a topological index directly on a gapped electronic band structure;
\item[\gls{ti}$_b$]
| necessarily possessing a conductive boundary bordering a trivially insulating material such as the vacuum;
\item[\gls{ti}$_x$]
| an insulating material $C$ for which an expansion $rC$ is conductive \cite{tqc-theory}.
\end{itemize}

An \emph{expansion} is defined within the third category, for a real scalar $r\ge1$, as a modified crystal $rC$, where for each atom at position $p$ in the original crystal $C$, the modified crystal $rC$ has a corresponding atom at $rp$. Thus, expansion increases the inter-atomic distances in the material. For $r\gg1$, an expansion $rC$ is regarded as forming an \emph{approximate vacuum}.

The relationships between the categories of \gls{ti} are displayed in Figure~\ref{fig:types_ti}.
Under certain circumstances, a \gls{ti} lying in one category may by implication ($\Rightarrow$) also lie in a second category. For example, the implication \gls{ti}$_e \Rightarrow$ \gls{ti}$_b$ forms a class of results known as \emph{bulk-boundary theorems} for specific topological indices \cite{topo-generic}. Further, \gls{ti}$_b \Rightarrow$ \gls{ti}$_x$ and \gls{ti}$_e \Rightarrow$ \gls{ti}$_x$ (Figure~\ref{fig:expansion}).
Finally, \gls{ti}$_b \Rightarrow$ \gls{ti}$_e$ is a trivial consequence of the fact that materials are specified by their electronic structure.

\begin{figure}
    \centering
\begin{tikzpicture}[
    node distance=4cm,
    auto,
    thick,
    ti/.style={circle, draw, text width=1cm, align=center}
]

\node[ti] (TIe) {TI$_e$};
\node[ti] (TIb) [right of=TIe] {TI$_b$};
\node[ti] (TIx) [below of=TIe, yshift=1cm] {TI$_x$};

\draw[->, dotted] (TIe) to [bend left=30] node[above] {Bulk-boundary correspondence} (TIb);
\draw[->] (TIb) -- (TIe);
\draw[->] (TIe) -- (TIx);
\draw[->] (TIb) -- (TIx);
\draw[->, dotted] (TIx) to [bend left=45] node[left=0.2cm] {\gls{tqc}} (TIe);


\end{tikzpicture}
    \caption{Logical relationships between different notions of \gls{ti}. 
    The dotted arrows represent relationships that may require additional assumptions to establish, reflecting the fact that, while bulk definitions of crystals have relatively simple translational symmetry, boundaries can be extremely complex.
    }
    \label{fig:types_ti}
\end{figure}


\begin{figure*}
    \centering
\begin{tikzpicture}[scale=1.4]

\draw[->] (2,0) -- (12,0) node[anchor=west] {$r$};

\draw (2,0.1) -- (2,-0.1) node[anchor=north] {TI ($r=1$)};
\draw (6,0.1) -- (6,-0.1) node[anchor=north] {Conductor};
\draw (10,0.1) -- (10,-0.1) node[anchor=north] {Mathematical Insulator ($r\rightarrow\infty$)};

\draw (1.25, 1.25) -- (2.75, 1.25);
\draw (1.25, 4.25) -- (2.75, 4.25);
\draw (1.25, 1.25) -- (1.25, 4.25);
\draw (2.75, 4.25) -- (2.75, 1.25);

\foreach \x in {1.5,2,2.5}
    \foreach \y in {1.5,2,2.5}
    {
        \fill[blue] (\x,\y) circle (0.1);
    }


\draw (5.25, 4.25) -- (6.75, 4.25);
\draw (5.25, 2.75) -- (5.25, 4.25);
\draw (6.75, 4.25) -- (6.75, 2.75);
\fill[draw=black,color=lightgray] (5.25, 2.75) rectangle (6.75, 1.0);

\foreach \x in {5.375,6,6.625}
    \foreach \y in {1.25,1.875,2.5}
    {
        \fill[blue] (\x,\y) circle (0.1);
    }

\draw (9.25, 2.75) -- (10.75, 2.75);
\draw (9.25, 4.25) -- (10.75, 4.25);
\draw (9.25, 2.75) -- (9.25, 4.25);
\draw (10.75, 4.25) -- (10.75, 2.75);

\foreach \x in {9,10,11}
    \foreach \y in {0.5,1.5,2.5}
    {
        \fill[blue] (\x,\y) circle (0.1);
    }

\foreach \x in {1.5,2,2.5}
    \foreach \y in {3,3.5,4}
    {
        \fill[blue] (\x,\y) circle (0.1);
    }

\foreach \x in {5.5,6,6.5}
    \foreach \y in {3,3.5,4}
    {
        \fill[blue] (\x,\y) circle (0.1);
    }

\foreach \x in {9.5,10,10.5}
    \foreach \y in {3,3.5,4}
    {
        \fill[blue] (\x,\y) circle (0.1);
    }


\filldraw[fill=gray!20] (1.5,5) rectangle (2.5,5.5); 
\filldraw[fill=gray!20] (1.5,6) rectangle (2.5,6.5); 
\draw[dotted, thick] (1.5,5.5) -- (2.5,6);
\draw[dotted, thick] (2.5,5.5) -- (1.5,6);

\filldraw[fill=gray!20] (5.5,5) rectangle (6.5,5.5); 
\filldraw[fill=gray!20] (5.5,6) rectangle (6.5,6.5); 
\draw[thick] (5.5,5.5) -- (6.5,6);
\draw[thick] (6.5,5.5) -- (5.5,6);

\filldraw[fill=gray!20] (9.5,5) rectangle (10.5,5.5); 
\filldraw[fill=gray!20] (9.5,6) rectangle (10.5,6.5); 

\node at (2,7) {$r=1$};
\node at (6,7) {$r=1.25$};
\node at (10,7) {$r\rightarrow\infty$};

\end{tikzpicture}
    \caption{
    An illustration of the expansion concept. For three different value ranges of the scalar $r$, crystal samples $C$ and $rC$ share a common interface. Above each respective physical picture (\gls{ti}, conductor and insulator), a schematic of the corresponding band structure for the $rC$ crystal is presented (valence bands below, conduction bands above). For $r=1$, where $rC=C$,  the conductive boundary surrounds both samples. For $r\gg1$, the expansion $rC$ forms an approximate vacuum, so the conductive border is around $C$. We locate a scalar $r'$ midway between the supremum of the set of $r$ such that $rC$ is a \gls{ti} and the infimum of the set of $r$ such that $rC$ is an approximate vacuum, marking this point with a notch in the diagram. If $C$ is a TI$_b$, consider the boundary states of the material $r'C$. Suppose that $r'C$ was insulating, i.e., the density of states falling within a certain energy range $[E_a, E_b]$ is $0$. Then, since expansions of insulating materials are insulating, the border of $r'C$ with both $C$ and the vacuum would be insulating. However, as the bordering $C$ is a \gls{ti}$_b$, at least one of the borders must be conductive, and so $r'C$ itself is conductive. Note the informality of this general argument, due to the difficulty of defining a \gls{ti}$_b$ directly. Nevertheless, for a \gls{ti}$_e$, the topological insulator status varies according to a continuous function of the electronic states, and therefore of $r$. In this case, $r'C$ is necessarily conductive. Thus, an \gls{ml} approach to \gls{tqc} classification must be sensitive to $r$ \cite{tqc-theory}.
    }
    \label{fig:expansion}
\end{figure*}

The prerequisites for the \gls{tqc} theory are group theory \cite{rotman2012introduction}, representation theory \cite{fulton2013representation, paxton2009introduction}, electronic structure \cite{kittel2018introduction}, and graph theory \cite{bollobas2013modern}. Comprehensive reviews exist \cite{cano2021band, cano2018building, van2022optical}. We generally follow the notation of the latter. \gls{tqc} utilizes the notion of an atomic limit with $r\gg1$ to establish a class of non-topological materials. These materials are used to generate a complete set of band representations. If a given band structure has symmetry indicators respecting the non-topological band representations, it is trivial. Otherwise it is topological. The \gls{tqc} algorithm may be extended to distinguish semimetals as well. However, it distinguishes topology for separated groups of bands, and is not exhaustive \cite{bouhon2020geometric, lange2021subdimensional}.

\section{Materials embeddings, machine learning architectures, and theoretical implementations}\label{s:design}

Previous research has demonstrated success with partial data models such as \gls{gbt}, random forests, $k$-nearest neighbor classifiers, support vector classifiers, and neural networks \cite{boateng2020basic}. Earlier \gls{gbt}s were successfully trained by data from \cite{tqc-learning, ml-shallow-resource}.

\begin{figure}
\centering
\begin{minipage}{.45\textwidth}
  \centering
\newcommand{\ElemLabel}[4]{
  \begin{minipage}{2.2cm}
    \centering
      {#2}%
      \linebreak \linebreak
      {\textbf{#3}}
  \end{minipage}
}

\newcommand{\Eleme}[4]{\ElemLabel{#1}{\Huge {#2}}{\Huge {#3}}{#4}}

\resizebox{!}{0.9 \textheight}{
\begin{tikzpicture}[font=\sffamily, rotate=90,transform shape]

  \tikzstyle{sblfill} = [fill=green!55]
  \tikzstyle{dblfill} = [fill=blue!25]
  \tikzstyle{pblfill} = [fill=orange!40]
  \tikzstyle{fblfill} = [fill=red!40]

  \tikzstyle{Element} = [minimum width=2.5cm, minimum height=2.5cm, node distance=2.75cm]
  \tikzstyle{dbl} = [Element, dblfill]
  \tikzstyle{pbl} = [Element, pblfill]
  \tikzstyle{sbl} = [Element, sblfill]
  \tikzstyle{fbl} = [Element, fblfill]
  \tikzstyle{PeriodLabel} = [font={\sffamily\LARGE}, node distance=2.0cm]
  \tikzstyle{GroupLabel} = [font={\sffamily\LARGE}, minimum width=2.75cm, node distance=2.0cm]

\node[sbl] (He) {\Eleme{2}{(0,0)}{He}{Helium}};
\node[below of=He, sbl] (Ne) {\Eleme{10}{(0,1)}{Ne}{Neon}};
\node[below of=Ne, sbl] (Ar) {\Eleme{18}{(0,2)}{Ar}{Argon}};
\node[below of=Ar, sbl] (Kr) {\Eleme{36}{(0,3)}{Kr}{Krypton}};
\node[below of=Kr, sbl] (Xe) {\Eleme{54}{(0,4)}{Xe}{Xenon}};
\node[below of=Xe, sbl] (Rn) {\Eleme{86}{(0,5)}{Rn}{Radon}};

\node[left of=He, sbl] (H) {\Eleme{1}{(1,0)}{H}{Hydrogen}};
\node[below of=H, sbl] (Li) {\Eleme{3}{(1,1)}{Li}{Lithium}};
\node[below of=Li, sbl] (Na) {\Eleme{11}{(1,2)}{Na}{Sodium}};
\node[below of=Na, sbl] (K) {\Eleme{19}{(1,3)}{K}{Potassium}};
\node[below of=K, sbl] (Rb) {\Eleme{37}{(1,4)}{Rb}{Rubidium}};
\node[below of=Rb, sbl] (Cs) {\Eleme{55}{(1,5)}{Cs}{Caesium}};

\node[left of=Li, pbl] (F) {\Eleme{9}{(2,1)}{F}{Fluorine}};
\node[below of=F, pbl] (Cl) {\Eleme{17}{(2,2)}{Cl}{Chlorine}};
\node[below of=Cl, pbl] (Br) {\Eleme{35}{(2,3)}{Br}{Bromine}};
\node[below of=Br, pbl] (I) {\Eleme{53}{(2,4)}{I}{Iodine}};
\node[below of=I, pbl] (At) {\Eleme{85}{(2,5)}{At}{Astatine}};

\node[left of=F, pbl] (O) {\Eleme{8}{(3,1)}{O}{Oxygen}};
\node[below of=O, pbl] (S) {\Eleme{16}{(3,2)}{S}{Sulphur}};
\node[below of=S, pbl] (Se) {\Eleme{34}{(3,3)}{Se}{Selenium}};
\node[below of=Se, pbl] (Te) {\Eleme{52}{(3,4)}{Te}{Tellurium}};
\node[below of=Te, pbl] (Po) {\Eleme{84}{(3,5)}{Po}{Polonium}};

\node[left of=O, pbl] (N) {\Eleme{7}{(4,1)}{N}{Nitrogen}};
\node[below of=N, pbl] (P) {\Eleme{15}{(4,2)}{P}{Phosphorus}};
\node[below of=P, pbl] (As) {\Eleme{33}{(4,3)}{As}{Arsenic}};
\node[below of=As, pbl] (Sb) {\Eleme{51}{(4,4)}{Sb}{Antimony}};
\node[below of=Sb, pbl] (Bi) {\Eleme{83}{(4,5)}{Bi}{Bismuth}};

\node[left of=N, pbl] (C) {\Eleme{6}{(5,1)}{C}{Carbon}};
\node[below of=C, pbl] (Si) {\Eleme{14}{(5,2)}{Si}{Silicon}};
\node[below of=Si, pbl] (Ge) {\Eleme{32}{(5,3)}{Ge}{Germanium}};
\node[below of=Ge, pbl] (Sn) {\Eleme{50}{(5,4)}{Sn}{Tin}};
\node[below of=Sn, pbl] (Pb) {\Eleme{82}{(5,5)}{Pb}{Lead}};

\node[left of=C, pbl] (B) {\Eleme{5}{(6,1)}{B}{Boron}};
\node[below of=B, pbl] (Al) {\Eleme{13}{(6,2)}{Al}{Aluminium}};
\node[below of=Al, pbl] (Ga) {\Eleme{31}{(6,3)}{Ga}{Gallium}};
\node[below of=Ga, pbl] (In) {\Eleme{49}{(6,4)}{In}{Indium}};
\node[below of=In, pbl] (Tl) {\Eleme{81}{(6,5)}{Tl}{Thallium}};

\node[left of=B, pbl] (Be) {\Eleme{4}{(7,1)}{Be}{Beryllium}};
\node[below of=Be, pbl] (Mg) {\Eleme{12}{(7,2)}{Mg}{Magnesium}};
\node[below of=Mg, pbl] (Ca) {\Eleme{20}{(7,3)}{Ca}{Calcium}};
\node[below of=Ca, pbl] (Sr) {\Eleme{38}{(7,4)}{Sr}{Strontium}};
\node[below of=Sr, pbl] (Ba) {\Eleme{56}{(7,5)}{Ba}{Barium}};

\node[left of=Ca, dbl] (Zn) {\Eleme{30}{(8,3)}{Zn}{Zinc}};
\node[below of=Zn, dbl] (Cd) {\Eleme{48}{(8,4)}{Cd}{Cadmium}};
\node[below of=Cd, dbl] (Hg) {\Eleme{80}{(8,5)}{Hg}{Mercury}};

\node[left of=Zn, dbl] (Cu) {\Eleme{29}{(9,3)}{Cu}{Copper}};
\node[below of=Cu, dbl] (Ag) {\Eleme{47}{(9,4)}{Ag}{Silver}};
\node[below of=Ag, dbl] (Au) {\Eleme{79}{(9,5)}{Au}{Gold}};

\node[left of=Cu, dbl] (Ni) {\Eleme{28}{(10,3)}{Ni}{Nickel}};
\node[below of=Ni, dbl] (Pd) {\Eleme{46}{(10,4)}{Pd}{Palladium}};
\node[below of=Pd, dbl] (Pt) {\Eleme{78}{(10,5)}{Pt}{Platinum}};

\node[left of=Ni, dbl] (Co) {\Eleme{27}{(11,3)}{Co}{Cobalt}};
\node[below of=Co, dbl] (Rh) {\Eleme{45}{(11,4)}{Rh}{Rhodium}};
\node[below of=Rh, dbl] (Ir) {\Eleme{77}{(11,5)}{Ir}{Iridium}};

\node[left of=Co, dbl] (Fe) {\Eleme{26}{(12,3)}{Fe}{Iron}};
\node[below of=Fe, dbl] (Ru) {\Eleme{44}{(12,4)}{Ru}{Ruthenium}};
\node[below of=Ru, dbl] (Os) {\Eleme{76}{(12,5)}{Os}{Osmium}};

\node[left of=Fe, dbl] (Mn) {\Eleme{25}{(13,3)}{Mn}{Manganese}};
\node[below of=Mn, dbl] (Tc) {\Eleme{43}{(13,4)}{Tc}{Technetium}};
\node[below of=Tc, dbl] (Re) {\Eleme{75}{(13,5)}{Re}{Rhenium}};

\node[left of=Mn, dbl] (Cr) {\Eleme{24}{(14,3)}{Cr}{Chromium}};
\node[below of=Cr, dbl] (Mo) {\Eleme{42}{(14,4)}{Mo}{Molybdenum}};
\node[below of=Mo, dbl] (W) {\Eleme{74}{(14,5)}{W}{Tungsten}};

\node[left of=Cr, dbl] (V) {\Eleme{23}{(15,3)}{V}{Vanadium}};
\node[below of=V, dbl] (Nb) {\Eleme{41}{(15,4)}{Nb}{Niobium}};
\node[below of=Nb, dbl] (Ta) {\Eleme{73}{(15,5)}{Ta}{Tantalum}};

\node[left of=V, dbl] (Ti) {\Eleme{22}{(16,3)}{Ti}{Titanium}};
\node[below of=Ti, dbl] (Zr) {\Eleme{40}{(16,4)}{Zr}{Zirconium}};
\node[below of=Zr, dbl] (Hf) {\Eleme{72}{(16,5)}{Hf}{Hafnium}};

\node[left of=Ti, dbl] (Sc) {\Eleme{21}{(17,3)}{Sc}{Scandium}};
\node[below of=Sc, dbl] (Y) {\Eleme{39}{(17,4)}{Y}{Yttrium}};
\node[below of=Y, dbl] (Lu) {\Eleme{71}{(17,5)}{Lu}{Lutetium}};

\node[left of=Lu, fbl] (Yb) {\Eleme{70}{(18,5)}{Yb}{Ytterbium}};
\node[left of=Yb, fbl] (Tm) {\Eleme{69}{(19,5)}{Tm}{Thulium}};
\node[left of=Tm, fbl] (Er) {\Eleme{68}{(20,5)}{Er}{Erbium}};
\node[left of=Er, fbl] (Ho) {\Eleme{67}{(21,5)}{Ho}{Holmium}};
\node[left of=Ho, fbl] (Dy) {\Eleme{66}{(22,5)}{Dy}{Dysprosium}};
\node[left of=Dy, fbl] (Tb) {\Eleme{65}{(23,5)}{Tb}{Terbium}};
\node[left of=Tb, fbl] (Gd) {\Eleme{64}{(24,5)}{Gd}{Gadolinium}};
\node[left of=Gd, fbl] (Eu) {\Eleme{63}{(25,5)}{Eu}{Europium}};
\node[left of=Eu, fbl] (Sm) {\Eleme{62}{(26,5)}{Sm}{Samarium}};
\node[left of=Sm, fbl] (Pm) {\Eleme{61}{(27,5)}{Pm}{Promethium}};
\node[left of=Pm, fbl] (Nd) {\Eleme{60}{(28,5)}{Nd}{Neodymium}};
\node[left of=Nd, fbl] (Pr) {\Eleme{59}{(29,5)}{Pr}{Praseodymium}};
\node[left of=Pr, fbl] (Ce) {\Eleme{58}{(30,5)}{Ce}{Cerium}};
\node[left of=Ce, fbl] (La) {\Eleme{57}{(31,5)}{La}{Lanthanum}};

  \node at (H.west -| Eu.north) [scale=2, font={\sffamily\Huge\bfseries}]
    {Left Step Periodic Table of Elements};

  \node[Element, scale=2] at (Ar.west -| Eu.north) {\Eleme{num}{(c, r)}{El}{Element}};

\end{tikzpicture}}
    \caption{The periodic table of elements, labeled with the column and row annotations used for \gls{ml}. Thus, the encoding for Gd is $(0,0,0,0,0,1,0;0,0,0,0,0,0,0,0,0,0,0,0,1,0,0,0;0)$ and O is $(0,1,0,0,0,0,0;0,1,0,0,0,0,0,0,0,0,0,0,0,0,0,0;1)$.}
    \label{fig:periodic_table}
\end{minipage}%
\hfil
\begin{minipage}{.45\textwidth}
  \centering
\resizebox{!}{0.85 \textheight}{\includegraphics{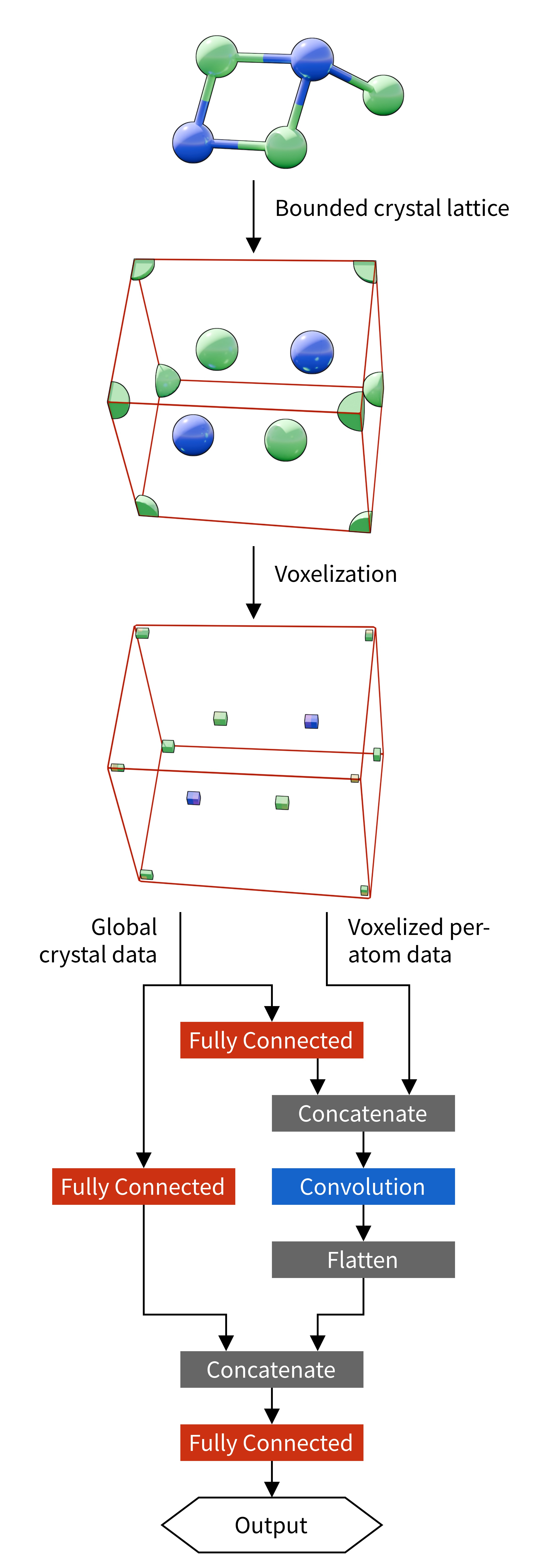}}
    \caption{The CCNN architecture using Gd$_2$O$_3$ as an example. A small cubic region surrounding one molecule of the crystal is converted into an antialiased voxel lattice. Each voxel encodes a user-configurable representation of an atom whose center is less than one voxel unit away from the voxel's center. Classification is performed by augmenting a fully connected network (red) with a series of convolution layers (blue) that process per-voxel atomic embeddings. }
    \label{fig:ccnn}
\end{minipage}
\end{figure}

Following training of the \gls{gbt} algorithm for topological data, a subsequent analysis demonstrated that electron counts and space groups were the primary distinguishing decision factors to determine material topology \cite{tqc-learning}. Model performance was excellent, peaking at 
$90\%$
for the full \gls{gbt} model. When the \gls{gbt} was coupled with \textit{ab initio} calculations that neglected spin-orbit coupling, accuracy peaked at $92\%$ on the materials with strong confidence in the predicted topological state. As full spin-orbit \textit{ab initio} calculations enable the direct prediction of material topology, these calculations were not used to supplement the \gls{ml} models. The primary benefit of using purely structure-based predictions is the encompassing generality, granting an easy method of retraining the models to new situations. 
Since the original dataset was not accessible,
the \gls{gbt} algorithm without \gls{dft} was exactly reconstructed, and applied to the current dataset. On the advanced \gls{tqc} dataset, it achieved an accuracy of 76\% as in Table~\ref{tab:ml_comparisonA}. All algorithms considered are compared to the $76\%$ benchmark, as no additional \textit{ab initio} calculations were included. In Ref.~\cite{tqc-learning}, the \gls{cgnn} was tested, but failed to converge to a reasonable accuracy for topological prediction. Now, it will be seen to have excellent predictive capability.

Four faithful embeddings of the underlying materials are tested. For each embedding, the data format is standardized as follows. Take $A$ to be the set of atoms in the primitive cell. Each atom $a\in A$ is associated with two types of information: the \emph{atomic identifier} $v_a$ and the \emph{atomic position} $p_a$. Finally, the \emph{global vector} $g$ is a vector containing primitive cell dimensions and symmetries. Different embeddings are considered for each of the input vectors, and tested over all \gls{ml} frameworks to determine the best representation.

For a classification with $n$ categories, recall that the \emph{one-hot} encoding of the $i$-th category is $0^{\oplus (i-1)} \oplus 1 \oplus 0^{\oplus (n-i)}$. To enhance generalization over merely using a one-hot embedding of atomic number, the embedding was chosen as $h(r)\oplus h(c \pmod 2) \oplus \lfloor c / 2 \rfloor$, using the left-step periodic table in Figure~\ref{fig:periodic_table} to supply $r$ and $c$ \cite{scerri2022various}. This allows generalization over the rows and columns of the periodic table with $7$ (rows) + $16$ (spinless columns) + $1$ (spin slot) = $24$ positions per atom. Embedding additional atomic properties was tested, but no additional performance gains were found.

The position embedding $p_a$ is network-dependent, but is stored using fractional units relative to the primitive cell basis. There are two major components of the global data vector $g$. The first gives the primitive cell dimensions using a sinusoidal encoding \cite{ml-fourfeat}, while the second records the space group with a one-hot embedding. Hyperparameter tuning was used solely for the \gls{tqc} dataset to demonstrate maximal network performance, and neglected for the remaining tests to demonstrate the ability to immediately generalize.

\subsubsection{Contemporary Naive Neural Network}

This approach employs a fully-connected feedforward neural network. To classify materials, the \gls{nn} maps a material's properties to a one-hot encoding of its classification.  Common to all materials in the datasets explored, there were fewer than $6$ of each type of atom in $A$. Thus, the atoms were partitioned by type into at most $6$ subsets and ordered from most common to least common as $A_1,A_2,\cdots ,A_6 \subseteq A$, respectively. Each subset $A_i$ has a corresponding maximum size $n_i$, and, as all $a \in A_i$ share the same $v_a$, the common atomic vector may be designated $v_i$. To account for when $|A_i| < n_i$, the empty position $p_{\emptyset}$ is set to $0^{\oplus |p_a|}$, the $0$-vector in the same vector space as $p_a$. Then, input to the \gls{nn} is organized into bins as $b_i = v_i \oplus \bigoplus_{a \in A_i} p_a^{\oplus |A_i|} \bigoplus p_{\emptyset}^{\oplus (n_i - |A_i|)}$, ensuring a fixed-size bin $b_i \in \mathbb{R}^{|v_i| + n_i}$ and therefore a constant-size input tensor for the \gls{nn}. Finally, all bins are concatenated as $g \oplus \bigoplus_{i=1}^6 b_i$.

\subsubsection{Current Crystal Graph Neural Networks}

\gls{cgnn}s are instances of convolutional graph neural networks applied to solid state materials \cite{xie2018crystal}. Information is embedded in each part of the graph. Here the global vector $g$ is considered as a vector separate from the graph. Each node is associated to $v_a$, and each edge $e = (a, b)$ has the information $v_e = (\overrightarrow{p_a - p_b}, |p_a - p_b|)$. During the graphical passes, the shape of each vector associated with the edges, vertices, and global data is maintained, allowing skip connections. In order to increase the descriptive capacity of the network, $v_a$, $v_e$ and $g$ are first embedded into the graph using networks $ NN_e^a, NN_e^e, NN_e^g$ to larger embedded vectors $v_e^v, v_e^e, v_e^g$. The final categorization is read from the last components of $v_e^g$. Thus, $v_e^g$ is at least the sum of the sizes of $g$ and the label vector.

\subsubsection{Novel Crystal Attention Neural Network}

A \gls{cann} is attention applied to encoded atoms. This generalizes deep set networks, which were previously found to exhibit extremely poor inference on materials datasets. Attention layers (notated as $MultiHead(Q, K, V)$ for query, key, and value matrices respectively) frequently operate on ordered structures \cite{vaswani2017attention}. However, attention naturally treats inputs as elements of a set. The \gls{cann} was designed based on the commonly known \emph{set transformer} framework applied to atoms  as described by \cite{lee2019set}. The equational status of the network is described by the input $Z = \bigoplus_{a \in A} (p_a \oplus v_a \oplus g)$ supplied to layers of feed-forward networks ($rFF$) as $rFF(SAB(MAB(SAB(Z))))$ with skip connections between every layer except the last. The layers are defined as  $\mathrm{MAB}(X, Y) = \mathrm{LayerNorm}(H + \mathrm{rFF}(H))$, where $H = \mathrm{LayerNorm}(X + \mathrm{MultiHead}(X, Y, Y))$. Then, augmented layers are defined as $\mathrm{SAB}(X) = \mathrm{MAB}(X, X)$, $\mathrm{ISAB}_m(X) = \mathrm{MAB}(X, MAB(I, X))$, and $\mathrm{PMA}_k(Z) = \mathrm{MAB}(S, \mathrm{rFF}(Z))$, where $I$ and $S$ are learned parameter sets.

This architecture allows for modeling pairwise and higher-order interactions among elements in the input set, while maintaining permutation invariance. The ISAB reduces the computational complexity from $O(n^2)$ to $O(nm)$, where $m$ is the number of inducing points, allowing the model to scale to large input sets. The set transformer is proven to be a universal approximator for permutation invariant functions, making it a powerful \gls{ml} tool for quantum materials modeling  \cite{lee2019set}.

\subsubsection{Innovative Crystal Convolutional Neural Network}

The final network examined is the \gls{ccnn}. This network uses a spatial representation of the atoms \cite{davariashtiyani2023voxel}. \gls{ccnn}s are instances of convolutional neural networks (CNN's) applied to solid state materials. Convolutional networks have been used extensively in both voxel and video domains, exploiting spatial and spatio-temporal uniformity by applying a kernel to a $2$-, $3$- or $4$-dimensional representation.

As visualized in Figure~\ref{fig:ccnn}, the tensorial embedding for the network is $N^3 \times (|v_{atom}|+1  +\nu_g)$ dimensional. The first three indices of the tensor are spatial indices, with the $N^3$ cube corresponding to the $[0,1)^3$ space consisting of the atoms' positions relative to the primitive cell spanning vectors. The addition of $\nu_g$ corresponds to generating an $N^3 \times \nu_g$ tensor directly from the global features via a multiperceptron network as $\nu_g = NN'(g)$. Tests demonstrated that concatenating $N^3 \times g$ to the voxel crystal cell was both computationally expensive, and failed to perform. To embed the atoms in the first $|v_{atom}|+1$ spots in the tensor, the atoms from the crystal are represented relative to the bounds of the 3D tensor using the relative coordinates in the crystal cell. Anti-aliasing is used to encode the atomic representations $v_a$ with a filling term directly into the voxel mesh \cite{zhang2019making}.

\section{Practical \gls{ml} implementations: results, applications, and discussion}\label{s:results}

The full models are capable of overfitting on any coherent set of training data to an arbitrary extent. Thus, training accuracy is not emphasized. For some tables, previous papers are used as approximate benchmarks for comparison. Since these papers may not use the same dataset, the comparisons are at best indicative.

One implication of the faithfulness of the models is that limits had to be introduced to speed training time. To compare a model to the \gls{gbt} algorithm, a penalty was assigned to the primitive cells unable to fit in the representation as follows: without knowledge of the underlying input variables, the best predictor of an element in the validation set $V$ is a single label, $p$, for each element of $V$. So, this optimal element was used as a default prediction for when materials were too large to use with the \gls{ml} models. Note that $p$ is extracted from the training set to prevent data contamination. For classification, $p$ is the most common label. For regression, if the loss is \gls{rmse} or \gls{mae}, then the $p$ which optimizes each of these measures of model error is the mean and median respectively. This gives a well-defined methodology to compare dissimilar models over an underlying dataset. 
It also gives a simple baseline model for comparisons, as 
represented in Tables 1, 2, and 3.

\subsection{General quantum materials property predictions}

The material representations are sufficient to determine the symmetry group. Thus, as a first test of the global power of the \gls{ml} algorithms, 151,000 materials were taken from the Materials Project and ICSD datasets \cite{mp, icsd}. The POSCAR file format \cite{vasp} was used as input to supply the \gls{ml} with atomic types and positions, and the primitive cell basis. The target variable for each material was the space group classification. The symmetry of a material is derived easily from the POSCAR description using structural geometry. Thus,  symmetry group classification is perfectly accurate, enabling a verification of the models' practical implementation.

Two primary implementations for the symmetry groups were tested: the one-hot encodings of the space and point groups. The space groups comprise $230$ labels, and the point groups comprise $32$ labels. As can be seen from Table~\ref{tab:symmetry-encodings}, \gls{ml} performance was low compared to analytic techniques. Indeed, this is a known weakness of \gls{ml}, and is an ongoing area of research in the \gls{ml} community. The \gls{ccnn} algorithm did manage to capture the majority of the space symmetries, indicating that spatial relationships are handled best with this direct approach, by comparison with the other three methods.

\begin{table}[h!]
    \centering
    \begin{tabular}{lcc}
        \toprule
        \textbf{Model} & \textbf{Point Group Accuracy} & \textbf{Space Group Accuracy} \\
        \midrule
        \hline
        \hline
        baseline & 0.15 & 0.10 \\
        \hline
        \gls{nnn} & 0.14 & 0.14 \\
        \hline
        \gls{cgnn} & 0.78 & 0.78 \\
        \hline
        \gls{ccnn} & 0.81 & 0.79 \\
        \hline
        \gls{cann} & 0.73 & 0.62 \\
        \bottomrule
    \end{tabular}
    \caption{Comparison of \gls{ml} models for space group and point group one-hot classification problems.}
    \label{tab:symmetry-encodings}
\end{table}

\textbf{}Formation energy per atom and the magnetic classification were both indexed from \cite{mp} for 151,000 materials. Natural errors were expected, due to temperature dependence for experimental results and 
limited
\gls{dft} accuracy. Performance on the magnetic dataset was strong, compared to the $81\%$ accuracy on a smaller dataset \cite{merker2022machine}. 
This illustrates the universality of model design as implemented for formation energy (Table~\ref{tab:ml_comparison}).
However, as expected, classification model performance for regression tasks without modification was weak,
which will be improved in the subsequent development.

\begin{table*}
    \centering
    \begin{tabular}{lcc}
        \toprule
        \textbf{Model} & \textbf{Formation energy MAE} & \textbf{Magnetic classification} \\
        \midrule
        \hline
        \hline
        Baseline comparison & 1.0 & 0.54 \\
        \hline
        \gls{nnn} & 0.26 & 0.75 \\
        \gls{cgnn} & 0.10 & 0.84 \\
        \gls{ccnn} & 0.19 & 0.79 \\
        \gls{cann} & 0.11 &  0.81 \\
        \bottomrule
    \end{tabular}
    \caption{Comparison of \gls{ml} models for the categorization problem. For MAE, a smaller number is better.}
    \label{tab:ml_comparison}
\end{table*}

\subsection{Topological classification}

Three primary sources were used to train the model. The first dataset \cite{tqc} contains a comprehensive list of topological indices for materials.
The material information was extracted in the form of POSCAR files from the two largest materials datasets available \cite{mp, icsd}. For each material, two sets of topological labels were extracted: $T_s$, a simplified labeling, and $T_r$, a refinement of $T_s$. Here, $T_s$ consists of three labels: LCEBR, TI, and SM, while $T_r$ consists of five augmented labels: LCEBR, NLC, SEBR, ES, and ESFD. There are $75,000$ materials with this labeling.

Two requirements were placed on the data. As a first criterion, primitive cells were required to have fewer than $60$ atoms. The second criterion arose from the issue that materials are often duplicated by stoichiometric label and symmetry group with minor variations in the POSCAR file. Thus, in cases where the topological labels agreed, the entries were condensed. In cases where there was a discrepancy in the topological data, the material was simply eliminated from the dataset, due to the high probability of a mistaken calculation, or of unusual ambient factors such as temperature and pressure. As an example of this type of situation, $39$ tuples of materials were merely minor distortions of each other, distinguished in the ICSD database, but identified in the Materials Project. After the filtration process, $36,580$ materials remained, with $455$ datapoints removed. The original dataset evidently contained thousands of duplicate materials. It is worth noting that an \gls{ml} process based on the original dataset would score artificially higher due to cross-contamination between the training and testing datasets. The topological composition of the dataset is ES, TI, SM, NLC, ESFD with 0.10, 0.27, 0.07, 0.07, and 0.49 as fractions of the whole dataset respectively.

\begin{table}[h!]
    \centering
    \begin{tabular}{lcc}
        \toprule
        \textbf{Model} & \textbf{Basic accuracy} & \textbf{Advanced accuracy} \\
        \midrule
        \hline
        \hline
        baseline & 0.49 & 0.49 \\
        GBT baseline\cite{tqc-learning} & 0.81 & 0.76 \\
        \hline
        \gls{nnn} & 0.72 & 0.65 \\
        \hline
        \gls{cgnn} & 0.83 & 0.80 \\
        \hline
        \gls{ccnn} & 0.76 & 0.71 \\
        \hline
        \gls{cann} & 0.77 & 0.72 \\
        \bottomrule
    \end{tabular}
    \caption{Comparison of \gls{ml} models for advanced \gls{tqc}.}
    \label{tab:ml_comparisonA}
\end{table}

The majority of model experiments were performed on the \gls{tqc} dataset. This enabled the diagnosis of specific model issues based on accuracy. Unless otherwise stated, all comments specifically pertain to the full 5 \gls{tqc} classifications. At the $49\%$ threshold, the model does not necessarily have information transfer between the input and the output, since the most common material type, non-topological, comprises $49\%$ of the dataset. An additional apparent plateau occurs near the $75\%$ accuracy range, after which training is diminished. The \gls{cgnn} model notably exceeds this threshold. Models were trained on the whole available dataset 20--60 times (epochs) to achieve maximal accuracy on the testing set. All four tested models exhibited an initial fast growth, then an apparent plateau that lasted for approximately one epoch before a more subtle long-term increase in accuracy became apparent. To account for dataset differences, an alternative \gls{gbt} algorithm was trained for Table~\ref{tab:ml_comparison}, based exactly on the specification provided in Ref.~\cite{tqc-learning} to compare approaches directly. All the models are either comparable to the \gls{gbt} baseline, or exceed it, as seen from Table~\ref{tab:ml_comparisonA}. 

The optimized implementation for each network is provided in \texttt{GitHub} 
with notes on optimization. Additional correlation effects were examined, showing weak correspondences between formation energy, magnetic classification, and topological effects in the supplementary material. Ensembles were created to test systemic model errors. Material misclassification was found to happen most frequently for less common elements, especially $\mathrm{Pt}$. Materials with multiple symmetries corresponding to the same stoichiometric formula were more frequently misclassified, unless the topological label was identical for all symmetry phases.

\section{Conclusion}\label{s:conclusion}

This work presents significant advancements in the application of \gls{ml} to predict, classify, and optimize the properties of quantum crystalline materials. By developing and refining multiple \gls{ml} architectures, including \gls{nnn}'s, \gls{cgnn}'s, \gls{cann}'s, and \gls{cgnn}'s, state-of-the-art performance is achieved on a variety of crucial quantum materials prediction problems. These models utilize the full representation of the crystal to diagnose difficult-to-predict materials with potentially novel quantum properties and physics. Additionally, the relative strengths and weaknesses of each model are cataloged for practical use.

The tools and models developed here are indexed online for public use and the simple development of new avenues for quantum materials prediction. The automated methods for data preprocessing, along with the full implementations and pretrained models provided on \texttt{GitHub}, can be efficiently applied to novel and pre-existing datasets. This enables the rapid classification and correlation of all crystalline materials, including quantum, magnetic, semiconducting, and topological properties.

\section*{Acknowledgment}

This work was supported as part of the Center for Energy Efficient Magnonics, an Energy Frontier Research Center funded by the U.S. Department of Energy, Office of Science, Basic Energy Sciences, under Award number DE-AC02-76SF00515.

\bibliography{citations}

\end{document}